# OPTICAL TRANSITION RADIATION (OTR) MEASUREMENTS OF AN INTENSE PULSED ELECTRON BEAM[†]

C. Vermare, D. C. Moir and G. J. Seitz
LANL, Los Alamos, N.M. 87454, USA

*Abstract*

We present the first time resolved OTR angular distribution measurements of an intense pulsed electron beam (1.7 kA, 60 ns). These initial experiments on the first axis of the **D**ual **A**xis **R**adiographic **H**ydro-**T**esting (DARHT) facility and subsequent analysis, demonstrate the possibility to extract, from the data, the energy and the divergence angle of a 3.8 and 20 MeV electrons.

## 1 INTRODUCTION

By using the OTR angular distribution property [1], it is possible to extract the energy and the angular dispersion of several kinds of beam [2]. Recently, Le Sage et al. [3] succeeded in producing a transverse phase-space mapping of a 100 MeV electron beam by coupling an interferometric measurement and a "mask" technique. We present the first OTR time resolved angular distribution measurements made on an intense pulsed electron beam with energies of 3.8 and 20 MeV.

The DARHT accelerator produces an intense pulsed electron beam (1.7 kA, 20 MeV, 60 ns) that impinges on a high-Z target. The quality of the X-ray source is determined by spot size and dose. The spot size is effected by magnet focal length, emittance, energy spread and beam motion. The dose is determined by beam energy, total charge, target material, and convergence angle. We present results of OTR angular distribution experiments performed at 3.8 and 20 MeV electron beam energies on the first axis of DARHT and compare these results to a 3D Ray-Tracing program which is able to calculate the effect of each electron beam parameter. Comparison shows that the data is most sensitive to the electron beam energy and the divergence/convergence angle. The maximum collection angle of the optical system limits results at low energy by mixing angular and spatial information.

The layout of the paper is as follows. The first section describes OTR angular distribution properties. This is followed by a brief description of the 3D Ray-Tracing program. The third section describes the experimental set-up. Section 4 shows initial OTR observations. Comparison between results and simulations are made in the fifth section. Time resolved measurements are described in section 6. Limitations and planned improvements of this diagnostic are discussed in the last section of this part.

## 2 OTR ANGULAR MEASUREMENTS

### 2.1 OTR properties

OTR is produced when a charged particle passes between media with different dielectric constants as a aluminium foil in vacuum. This light is emitted with a characteristic angular distribution that depends of the particle energy and direction. The Fig. 1 shows the OTR density versus angle on the incidence plane (plane defined by the beam axis and the normal vector of the target). The zero angle corresponds to the "specular" direction.

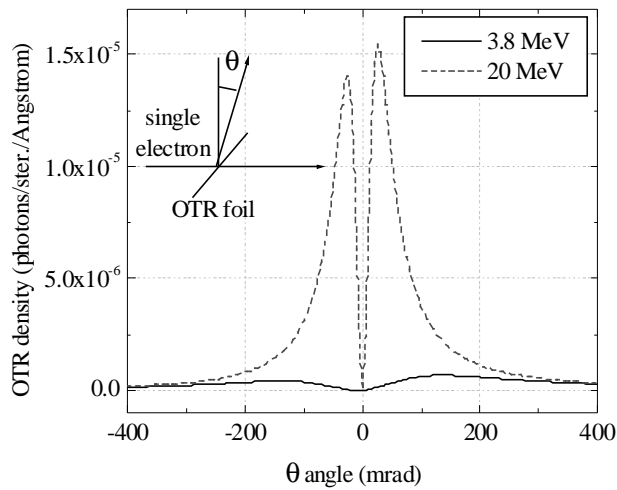

**Figure 1: OTR density versus angle on the incidence plane for two different energy (the tilt angle of the foil is 45 degree).**

It is important to note that the light produced by OTR is distributed with two different polarizations.

### 2.2 Ray-Tracing program

The system composed of the beam, the OTR foil and the detection system has 3D geometry. We developed a Ray-Tracing program able to follow photons produced by each electron of a phase-space-defined beam. The code calculates the image received by the screen for each polarisation. For a simulation, a spectrum range, foil material, lens and screen size and position and the tilt angle of the target are chosen. The results obtained are absolute as shown in Fig. 1.

---

## 2.3 Experimental set-up

The experimental set-up is shown in Fig. 2.

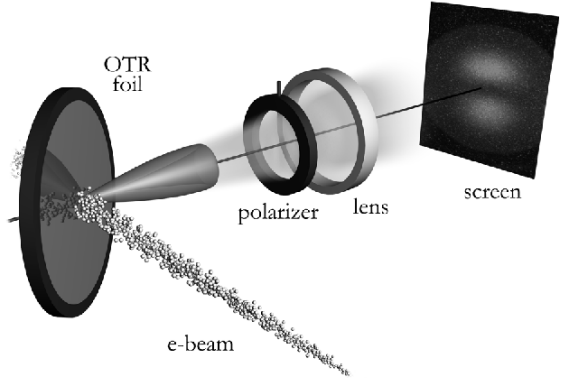

**Figure 2: Experimental set-up for angular distribution measurement.**

The electron beam passes through a thin aluminised Kapton foil (10 microns) and produces OTR photons from the aluminium-vacuum interface. This target is tilted 45 degrees so the center of the OTR distribution is emitted at 90 degrees from the beam axis. The OTR light is collected by an achromatic doublet (focal length = 200 mm and effective diameter = 70 mm) which produces an angular image of the source on a screen. This geometry requires the distance between the lens and the screen to be equal to the focal length. The distance between the target and the lens determine the maximum collection angle of the system. Limitations are the screen size, the "Cherenkov" background created inside the lens by secondary electrons or X-rays and discoloration of the optics caused by radiation damage. A 200-mm focal length achromatic lens that gives a maximum collection angle of 170 mrad is used. A polarizer can be added between the lens and the screen to separate each polarisation. An 8-frame gated camera records the image formed on the screen. This device splits the light up to 8 different micro-channel plate (MCP) gated CCD. The gating system of each MCP makes it possible to record each image with a 10 nanoseconds duration.

This optical system is to first order independent of the beam size and position. However, with large angle of the OTR lobe (specially at low energy), a large beam size can affect the angular distribution. In this case, a precise analysis with the Ray-Tracing program is required.

## 2.4 OTR confirmation

To confirm that OTR is the main part of the light collected by the system as opposed to prompt Cherenkov light generated by X-rays or secondary electrons, initial measurements were made of the polarization of light from the source. The Fig. 3 shows three images corresponding to (a) both polarizations (no polarizer), (b) polarization in the incidence plane and (c) polarization in the observation plane.

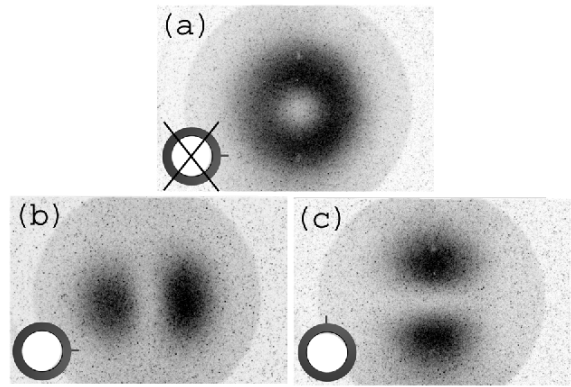

**Figure 3: Experimental picture of the OTR polarisation. The beam energy is 3.8 MeV**

These pictures demonstrate that the light observed is OTR. A second test was to block the light coming from the target by placing a thin aluminium foil between the target and the lens. The result confirms the low ratio of the "Cherenkov" light coming from the target. Tests were made at both 3.8 MeV and 20 MeV.

## 2.5 Extraction of beam parameters

According to the 3D model of OTR, a horizontal (vertical) cut of the picture recorded gives information about the beam. Results indicate that the distance between the maximum positions is determined by the energy and the filling at the centre is most sensitive to beam envelope divergence/convergence angle at the detector.

By changing the current on the guiding solenoid before the OTR foil, the beam size and divergence angle are varied. The Fig. 4 shows the effect predicted by the simulations. These curves correspond to the full beam duration and they are normalised.

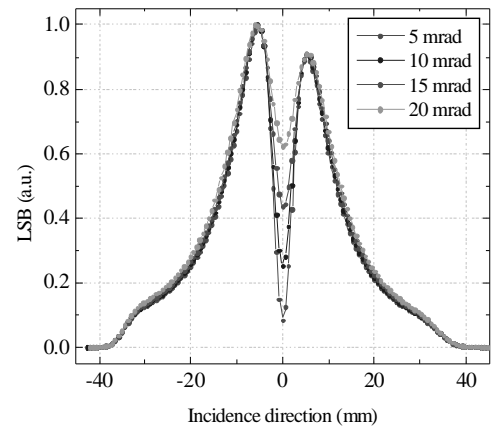

**Figure 4: Simulations of the effect of the divergence angle on the OTR angular distribution (20 MeV).**

Fig. 5 shows a comparison of experimental data with convergence angle determined by the magnet setting demonstrating the sensitivity observed in the calculations.

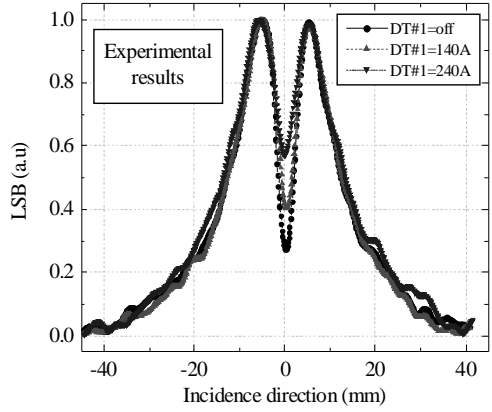

**Figure 5: Experimental results of the effect of the divergence angle on the OTR angular distribution (20 MeV).**

For the energy measurement, induction cells were de-energized at the end of the accelerator. An energy range between 15 MeV and 20 MeV was obtained. Fig. 6 shows the overlaid experimental results.

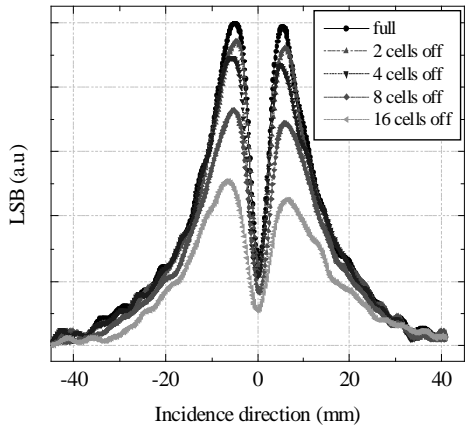

**Figure 6: Experimental results of the effect of the beam energy on the OTR angular distribution.**

These results are in agreement with the simulations. The precision on the energy measurement for these data is 250 keV at 20 MeV and 150 keV at 15 MeV.

### 2.6 Time resolved measurement

Fig. 7 is an example of a time resolved result. The beam energy is constant through the pulse. Therefore, we can observe the beam divergence/convergence as a function of time. For this measurement, the vacuum was reduced to $3.10^{-5}$ Torr to initiate a time dependent focusing of the beam due to background gas neutralization. This effect is, also, observed in the spatial measurement. This effect disappears then the vacuum reach $5.10^{-6}$ Torr. These results need to be improved. Currently, they show a time variation of the divergence angle about 5 mrad.

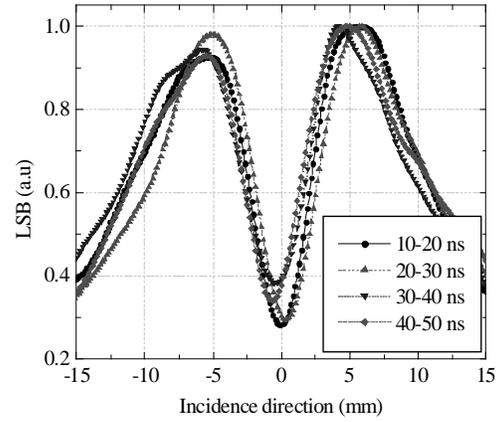

**Figure 7: Time resolved measurement of the beam parameter using the OTR angular distribution (20 MeV).**

### 2.7 Discussion and perspectives

If the system is carefully aligned and the maximum angle collected is more than four times $1/\gamma$, each beam parameter can be consider independent. If not, they are linked and a direct comparison with different simulations is necessary. We found this complication at 3.8 MeV with our set-up.

After these measurements, we plan different amelioration. First of all, the CCD dynamic range must be improved (256 to 64k levels). There are calibration concerns associated with the 8 different CCD camera that need to be addressed before we can extract quantitative information from the images. Next, we will use a streak camera image a slice of the OTR distribution directly as a function of time. Also, the magnification of the OTR angular distribution image must be increase to obtain a better definition of the maximum and the centre.

## CONCLUSION

The results presented here prove that it is possible to use the OTR angular distribution information to measure the energy and divergence/convergence angle of an intense pulsed electron beam. The light intensity is sufficient for time-resolved measurement of these parameters in the 3.8-20 MeV energy range.

## REFERENCES

[1] **L. Wartski**, "Etude du rayonnement de transition optique produit par des electrons de 30 a 70 MeV", Thesis, Orsay (1976).
[2] **D. W. Rule**, "Transition radiation diagnostics for intense charged particle beams", Nucl. Inst. Meth., B24/25 (1987) 901-904..
[3] **G.P. Le Sage et al.**, "Transverse phase space mapping of a relativistic electron beam using Optical Transition Radiation", PRST-AB, 2 (1999), 122802.